\documentclass[sigconf,nonacm]{acmart}

\usepackage{xspace}
\usepackage{amsmath,amsfonts}
\usepackage{algorithmic}
\usepackage{algorithm}
\usepackage{array}
\usepackage[caption=false,font=normalsize,labelfont=sf,textfont=sf]{subfig}
\usepackage{pgfplots}
\usepgfplotslibrary{units} 
\pgfplotsset{compat=newest} 
\usepackage{textcomp}
\usepackage{stfloats}
\usepackage{url}
\usepackage{verbatim}
\usepackage{graphicx}
\usepackage{multirow}
\usepackage{booktabs}
\usepackage{nameref}
\usepackage{colortbl}
\usepackage{tabularx}
\usepackage{multirow}
\usepackage{lipsum}
\usepackage{hyperref}
\usepackage{mathtools}
\usepackage{xcolor}
\usepackage{soul}
\setul{0.1ex}{0.5ex}
\usepackage[most]{tcolorbox}
\usepackage{dsfont}
\hypersetup{
    colorlinks=true,
    linkcolor=black,
    filecolor=black,      
    urlcolor=black,
    citecolor=black,
    }

\usepackage{siunitx}

\usepackage{tikz}
\usetikzlibrary{positioning}
\usepackage{tikz}
\usepackage{float}
\usetikzlibrary{shapes.geometric, arrows, chains}
\tikzstyle{main} = [rectangle, minimum width=1.5cm, minimum height=0.6cm,text centered, text width=2cm, draw=black]
\tikzstyle{arrow} = [thick,->,>=stealth]
\tikzstyle{l} = [draw, -latex',thick]

\definecolor{A1}{HTML}{feeba2}
\definecolor{A2}{HTML}{fed069}
\definecolor{A3}{HTML}{feaa38}
\definecolor{A4}{HTML}{f3801c}
\definecolor{A5}{HTML}{d85a09}

\AtBeginDocument{%
  }

\setcopyright{none}
\copyrightyear{2024}
\acmYear{2024}

\acmConference{}

\newcommand{\tool}{\textsc{SecRush}\xspace}
\newcommand{\metric}{\textsc{SecScore}\xspace}

\begin{document}

\title{\metric: Enhancing the CVSS Threat Metric Group with Empirical Evidences}

\author{Miguel Santana}
\affiliation{%
  \institution{Banco de Portugal}\country{Portugal}
}
\author{Vinicius V. Cogo}
\affiliation{%
  \institution{LASIGE, Informática, Faculdade de Ciências, Universidade de Lisboa}\country{Portugal}
}
\author{Alan Oliveira de Sá}
\affiliation{%
  \institution{LASIGE, Informática, Faculdade de Ciências, Universidade de Lisboa}\country{Portugal}
}

\renewcommand{\shortauthors}{}
\renewcommand{\shorttitle}{}
\settopmatter{printfolios=true}

\begin{abstract}
\textbf{\emph{Background:}} Timely prioritising and remediating vulnerabilities are paramount in the dynamic cybersecurity field, and one of the most widely used vulnerability scoring systems (CVSS) does not address the increasing likelihood of emerging an exploit code.\\ 
\textbf{\emph{Aims:}} We present \mbox{\metric}, an innovative vulnerability severity score that enhances CVSS Threat metric group with statistical models from empirical evidences of real-world exploit codes.\\
\textbf{\emph{Method:}} \metric adjusts the traditional CVSS score using an explainable and empirical method that more accurately and promptly captures the dynamics of exploit code development. \\
\textbf{\emph{Results:}} Our approach can integrate seamlessly into the assessment /prioritisation stage of several vulnerability management processes, improving the effectiveness of prioritisation and ensuring timely remediation. 
We provide real-world statistical analysis and models for a wide range of vulnerability types and platforms, demonstrating that \metric is flexible according to the vulnerability's profile. 
Comprehensive experiments validate the value and timeliness of \metric in vulnerability prioritisation. \\
\textbf{\emph{Conclusions:}} \metric advances the vulnerability metrics theory and enhances organisational cybersecurity with practical insights.
\end{abstract}

\settopmatter{printacmref=false}
\settopmatter{authorsperrow=3}
\maketitle


\section{Introduction}

\begin{figure*}[b!]
\centering\includegraphics[width=1\textwidth]{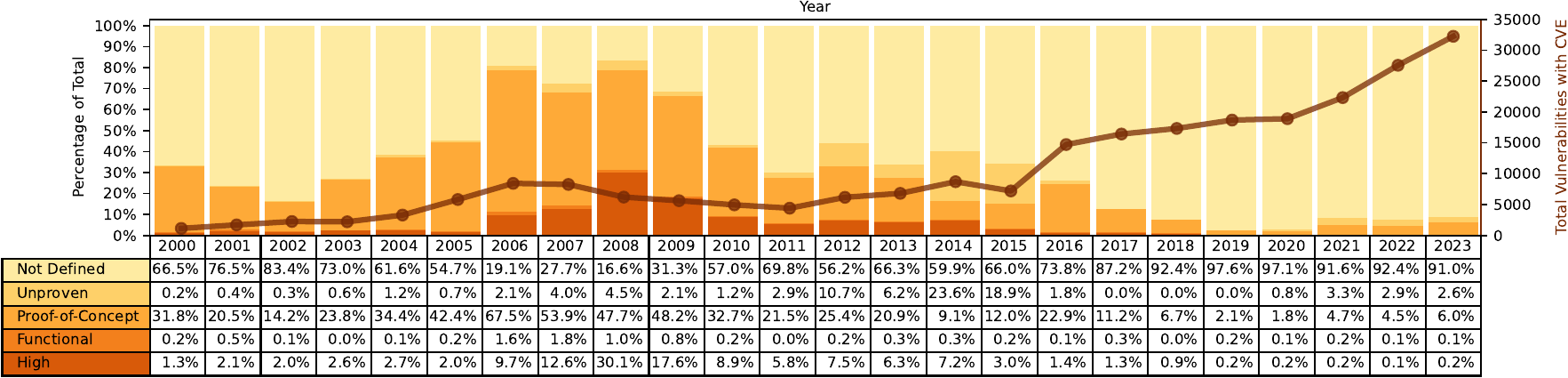}
\vspace{-6.5mm}
\caption{Statistics on the yearly published vulnerabilities and exploit maturity in VulDB~\cite{vuldb} from $2000$ to $2023$.}
\label{fig:vuldb_statistics}

\end{figure*}

Vulnerabilities are flaws that attackers can exploit in attempts to intrude a system, leading to security failures if protective measures fail to refrain them~\cite{10.1007/3-540-45177-3_1}.
Vulnerability prevention and removal are protective measures essential in safeguarding a system against potential exploits and attacks.
Organisations usually leverage vulnerability management processes to detect, prioritise, remediate, and mitigate vulnerabilities, whose automation directly propels their practicality~\cite{foreman2019vulnerability}.


Several reports acknowledge the worrisome growth in the number of novel identified vulnerabilities along the years~\cite{Greig2021, InformationTechnologyLaboratory2021}.
For instance, the brown line in Figure~\ref{fig:vuldb_statistics} displays the yearly number of published vulnerabilities in VulDB~\cite{vuldb}, which roughly grown from approximately $1$k in 2000 to $5$k in 2010 and to $19$k in 2020.
The unbounded nature of this growth becomes even more evident as $27$k vulnerabilities were published in 2022 and $32$k in 2023.

Vulnerability management processes rely upon widely-adopted vulnerability identifiers (e.g., CVE ID~\cite{cve}) and scores (e.g., CVSS~\cite{cvss}, EPSS~\cite{epss}) to timely prioritise that many vulnerabilities.
CVSS~\cite{cvss} (currently at version~$v4$) stands as the most prominent vulnerability score, featuring a Base component (called \mbox{CVSS-B}) that serves mainly as a technical severity assessment for vulnerabilities---\emph{i.e., it does not fit for standalone risk evaluation}~\cite{cvss4}. 

Integrating this Base score with Threat and Environment metric groups composes the \mbox{CVSS-BTE}~\cite{cvss4}, a more practical, adaptive vulnerability score variant.
The Threat metrics group adjusts the CVSS Base score based on the Exploit Maturity as a means of estimating an increased (resp. decreased) likelihood of a vulnerability being attacked due to the availability of a more (resp. less) mature exploit technique or code (e.g., CVSS~v3.1~\cite{cvss} categorises it in ``\texttt{Not Defined}'', ``\texttt{Unproven}'', ``\texttt{Proof-of-Concept}'', ``\texttt{Functional}'', and ``\texttt{High}'').
Problems in this approach include, but are not limited to, the fact that \emph{CVSS saturates when the maturity of an exploit code is ``\texttt{Not Defined}'' and these status are manually changed by experts (i.e., reactively, not time-dependent), taking years to see them reflected in vulnerability databases}.

This reality is illustrated also in Figure~\ref{fig:vuldb_statistics} through the stacked bars and table indicating that it takes seven years for the ``\texttt{Not Defined}'' category to comprise less than $90\%$ of the yearly published vulnerabilities in VulDB~\cite{vuldb} and fifteen years to comprise less than $50\%$.
Furthermore,  the ``\texttt{High}'' and ``\texttt{Functional}'' categories have never collectively represented more than one-third of the yearly published vulnerabilities.
Nevertheless, \emph{prioritising vulnerabilities clearly requires a timelier approach to effectively become practical}.

The Exploit Prediction Scoring System (EPSS)~\cite{epss} is a machine learning (ML) approach for estimating the probability of a software vulnerability being exploited within the following 30 days.
It complements CVSS since it produces a data-driven score between 0 and~1 that is directly proportional to the likelihood of exploitation of a given vulnerability in the wild. 
Although EPSS has been improving its accuracy (from an F1-score of $0.299$ in EPSS~v1 to $0.451$ in v2, and $0.728$ in v3~\cite{epss}), \emph{it still requires better AI interpretability and seamless integration with CVSS, as well as it must be recalculated for all known vulnerabilities during each prioritisation round}.

In this work, we introduce \metric, an innovative, empirical scoring system that enhances the CVSS Threat metric group by directly integrating an explainable metric, thereby enabling an effective vulnerability prioritisation. 
This score is adeptly designed to be incorporated not only within our proposed vulnerability management process, \tool, but also alongside other existing prioritisation methods, such as EPSS~\cite{epss}, to ensure comprehensive and timely remediation of vulnerabilities.
In summary, the paper contributions are:
\begin{enumerate}
    \item An empirical study into the likelihood of existing exploits for a wide range of vulnerability profiles (\S\ref{sec:secscore_statistics});
    \item The formulation of \metric, a vulnerability severity score that integrates the knowledge from our empirical study into the CVSS Threat metric group (\S\ref{subsec:secscore});
    \item An experimental evaluation of \metric, demonstrating its value and timeliness in vulnerability prioritisation (\S\ref{sec:impact_prioritisation}).
\end{enumerate}

\section{Context and Background} \label{sec:bkg}

This section contextualises vulnerability management processes, illustrating where one can integrate \metric, and provides an overview of the Common Vulnerability Scoring System (CVSS) as the background for the modifications proposed in \metric.


\subsection{Vulnerability Management Processes} \label{subsec:vuln-mgmt-proc}

Vulnerability management processes (VMPs) arrange vulnerability management tasks into, typically, four to six consecutive stages according to the organisational methodologies and technologies.
In general, VMPs include \setulcolor{A1}\ul{discovery/scanning}, \setulcolor{A2}\ul{assessment/prioritisation}, \setulcolor{A3}\ul{report}, \setulcolor{A4}\ul{remediation}, \setulcolor{A5}\ul{verification/validation}, and \setulcolor{A3}\ul{retrospective} stages.
Table~\ref{tab:vmp-stages} compares the stages of several VMPs~\cite{Humphries2020,Rapid7,TeamAscend2019,FordhamUniversity2021,Tenable2020b,UniversityofMiami,Cavalancia2020,CenterforDiseaseControlandPrevention2021}.
They differ on the number and naming of stages, the automation of the discovery, the assessment and prioritisation order, the position of the report stage (e.g., before or after remediation), and the existence or absence of verification/validation.

\setlength\tabcolsep{5 pt}
\begin{table}[t!]
\begin{center}
\caption{The stages of vulnerability management processes.}
\vspace{-3mm}
\label{tab:vmp-stages}
\begin{tabularx}{\columnwidth}{l|l||XlXXXX}
\hline
\textbf{VMP} & \textbf{\#} & \multicolumn{6}{c}{\textbf{Stages}} \\ \hline \hline

Exabeam~\cite{Humphries2020}&\multirow{2}{*}{4}&\cellcolor{A1}\emph{I}&\cellcolor{A2}\emph{E}&\cellcolor{A2}&\cellcolor{A4}\emph{Rem.}&\cellcolor{A4}&\cellcolor{A3}\emph{Rep.}\\ \cline{1-1}\cline{3-8}

Rapid7~\cite{Rapid7}& &\cellcolor{A1}\emph{S}&\cellcolor{A2}\emph{A}&\cellcolor{A2}&\cellcolor{A4}\emph{Add.}&\cellcolor{A4}&\cellcolor{A3}\emph{Rep.$^*$}\\ \hline

Ascend~\cite{TeamAscend2019}&\multirow{4}{*}{5}&\cellcolor{A1}\emph{D}&\cellcolor{A2}\emph{A}&\cellcolor{A3}\emph{Rep.}&\cellcolor{A4}\emph{Rem.}&\cellcolor{A5}\emph{Ver.}&\cellcolor{A5}\\	 \cline{1-1}\cline{3-8}

Fordham U.~\cite{FordhamUniversity2021}& &\cellcolor{A1}\emph{D}&\cellcolor{A2}\emph{Pri.}~ ~\emph{Pl.} &\cellcolor{A2}&\cellcolor{A4}\emph{Rem.}&\cellcolor{A5}\emph{Val.}&\cellcolor{A5}\\ \cline{1-1}\cline{3-8}

Tenable~\cite{Tenable2020b}& &\cellcolor{A1}\emph{D}&\cellcolor{A2}\emph{A}~ ~ ~\emph{Pri.}&\cellcolor{A2}&\cellcolor{A4}\emph{Rem.}&\cellcolor{A5}\emph{M}&\cellcolor{A5}\\	 \cline{1-1}\cline{3-8}

U. Miami~\cite{UniversityofMiami}& &\cellcolor{A1}\emph{P}~ ~\emph{S}&\cellcolor{A2}\emph{Def.}&\cellcolor{A2}&\cellcolor{A4}\emph{Imp.}&\cellcolor{A5}\emph{RS}&\cellcolor{A5}\\ \hline

AT\&T~\cite{Cavalancia2020}&\multirow{2}{*}{6}&\cellcolor{A1}\emph{D}&\cellcolor{A2}\emph{Pri.}~ ~ ~\emph{A}&\cellcolor{A2}&\cellcolor{A4}\emph{Rem.}&\cellcolor{A5}\emph{Ver.}&\cellcolor{A3}\emph{Rep.}\\ \cline{1-1}\cline{3-8}

CDC~\cite{CenterforDiseaseControlandPrevention2021}& &\cellcolor{A1}\emph{D}&\cellcolor{A2}\emph{Pri.}~ ~ ~\emph{A}&\cellcolor{A3}\emph{Rep.}&\cellcolor{A4}\emph{Rem.}&\cellcolor{A5}\emph{Ver.}&\cellcolor{A5} \\ \hline


\end{tabularx}
\vspace{-1mm}
\end{center}
\emph{
\hspace{-0.5mm}\#~=~Num. of Stages, \colorbox{A2}{A}=~Assess, \colorbox{A4}{Add.}=~Pri.\&Address, \colorbox{A1}{D}=~Discover, \colorbox{A2}{Def.}=~Define~Rem., \colorbox{A2}{E}=~Evaluate, \colorbox{A1}{I}=~Identify, \colorbox{A4}{Imp.}=~Implement Rem., \colorbox{A5}{M}=~Measure, \colorbox{A1}{P}=~Prepare, \colorbox{A2}{Pl.}=~Plan, \colorbox{A2}{Pri.}=~Prioritise, \colorbox{A4}{Rem.}=~Remediate, \colorbox{A3}{Rep.}=~Report, \colorbox{A5}{RS}=~Rescan, \colorbox{A1}{S}=~Scan, \colorbox{A5}{Val.}=~Validate, \colorbox{A5}{Ver.}=~Verify, \colorbox{A3}{$^*$}~Rep. and Cont. Management.}
\vspace{-4mm}
\end{table}


    
    
    

The \setulcolor{A1}\ul{discovery/scanning} stage identifies vulnerabilities and associates them with manually, systematically, or continuously collected catalogues of organisation's physical and virtual assets. 
The \setulcolor{A2}\ul{assessment/prioritisation} stage ranks the identified vulnerabilities based on their risk to the organisation.
Many VMPs already obtain risk scores for the identified vulnerabilities from publicly-available open catalogues (e.g., the CVE by MITRE~\cite{cve}).
However, the staggering proportion of vulnerabilities classified at the highest risk level ultimately discredits the current vulnerability risk assessment process.
We propose \metric, an innovative vulnerability severity score for this VMP stage considering statistical models built with empirical evidences from real-world exploit codes.

Usually, the \setulcolor{A3}\ul{report} stage documents the prioritised vulnerabilities to guide resolution efforts with detailed, actionable mitigation plans.
During the \setulcolor{A4}\ul{remediation} stage, vulnerabilities are mitigated based on their severity and availability of patches.
Misconfigurations must be corrected and unpatched vulnerabilities must be continuously monitored to prevent escalation.
If included in the VMP, the \setulcolor{A5}\ul{verification/validation} stage assesses the mitigation effectiveness in reducing the organisation exposure to the prioritised vulnerabilities.
Cybersecurity teams may conduct standard routines or opt for on-demand vulnerability scans to ensure that vulnerabilities were properly managed.
Finally, VMPs that include a \setulcolor{A3}\ul{retrospective} stage usually promote reflective meetings to review the overall VMP effectiveness, periodically reporting the resolved vulnerabilities and providing insights into the ongoing risk management process.

Four-stage VMPs (e.g.,~\cite{Humphries2020,Rapid7}) are often considered a simplistic approach that overlooks the cumulative impact of fixing all vulnerabilities, fails to validate the risk reduction, neglects the presence of exploits in the evaluation phase, and lacks continuous vulnerability identification.
Five-stage VMPs (e.g.,~\cite{TeamAscend2019,FordhamUniversity2021,Tenable2020b,UniversityofMiami}) provide a more holistic approach that confirms vulnerability mitigation and formally prioritises vulnerabilities.
Six-stage VMPs (e.g.,~\cite{Cavalancia2020,CenterforDiseaseControlandPrevention2021}) focus on continuous risk evaluation
and automated discovery, although challenges remain in monitoring vulnerabilities without patches and ensuring an effective prioritisation.

\subsection{Common Vulnerability Scoring System}
\label{subsec:cvss}

Vulnerability prioritisation challenges are frequently discussed in the literature, highlighted by the various solutions based on diverse vulnerability scoring systems.
The Common Vulnerability Scoring System (CVSS)~\cite{cvss}, managed by the Forum of Incident Response and Security Teams (FIRST), is a prominent framework for assessing vulnerability severity.
Although CVSS should not determine alone vulnerability prioritisation~\cite{cvss,cvss4,spring2021time}, it still serves as a foundational metric for many prioritisation methodologies.

Introduced in 2005~\cite{cvss1}, CVSS has undergone several updates. 
As of the end of 2023, CVSS$_{v4}$~\cite{cvss4} has been officially adopted, though many publicly-available vulnerability databases continue to use its predecessor, CVSS$_{v3.1}$~\cite{cvss31}. 
For this reason, the present section details CVSS$_{v3.1}$. 
Notwithstanding, \metric is designed to be compatible with both CVSS$_{v3.1}$ and CVSS$_{v4}$ (see Figures~\ref{fig:forensics} and~\ref{fig:overview_cvss4}), ensuring broad applicability across different versions as organisations transition to the latest standard.
Figure~\ref{fig:overview_cvss31} provides an overview of the CVSS$_{v3.1}$ equations and metrics, which will be described next.
The CVSS$_{v3.1}$ score of a vulnerability is computed considering three components: a Base Score~$\mathcal{B}$; a Temporal Score~$\mathcal{T}$; and an Environmental Score~$\mathcal{E}$. 



The CVSS$_{v3.1}$ Base Score~$\mathcal{B}$ is calculated using the impact~$\mathcal{I}$ and exploitability~$\mathcal{X}$ of the vulnerability.
Impact~$\mathcal{I}$ is derived from Equation~\ref{eq:cvss-base-impact}, involving the Impact Sub-Score~$\mathit{ISS}$ and the scope metric~$S$.
$\mathit{ISS}$ assesses the vulnerability's effects on the Confidentiality~$C$, Integrity~$I$, and Availability~$A$ of an asset.
Scope~$S$ determines whether the vulnerability's impact is confined withing the same security authority ($S = \overline{Changed}$) or extends beyond it ($S = Changed$), affecting additional resources.
Exploitability~$\mathcal{X}$ is determined by Equation~\ref{eq:cvss-base-exploitability}, which incorporates the base metrics of Attack Vector~($AV$), Attack Complexity~($AC$), Privileges Required~($PR$), and User Interaction~($UI$). 
The Base Score \(\mathcal{B}\) is then calculated using Equation~\ref{eq:cvss-base}, which includes a roundup operation~$\mathcal{R}$ to the nearest tenth, ensuring the result is the smallest decimal equal to or greater than the input, as detailed in Equation~\ref{eq:round}.

\begin{figure}[t!]
\centering
\begin{tcolorbox}[size=small,colback=A1,colframe=A3,title=,subtitle style={boxrule=0.4pt,colback=A3,colupper=black}]
\footnotesize
\vspace{-1mm}
\tcbsubtitle{\textbf{CVSS$_{v3.1}$ Scores and Metrics:} }

\medskip

\begin{tabularx}{\columnwidth}{ccX}
\hline
  \parbox[t]{2mm}{\multirow{9}{*}{\rotatebox[origin=c]{90}{\textbf{Base Score ($\mathcal{B}$)}}}}  &  & Attack Vector ($\mathit{AV}$)       \\
    & \textbf{Exploitability}  & Attack Complexity ($\mathit{AC}$)   \\
    & \textbf{Score ($\mathcal{X}$)}  & Privileges Required ($\mathit{PR}$) \\
    & & User Interaction ($\mathit{UI}$)    \\ \cline{2-3}
    & \textbf{Scope}  & \multirow{2}{*}{Scope ($S$)}  \\ 
    & \textbf{Score ($\mathcal{S}$)} &   \\ \cline{2-3}
    & \textbf{\multirow{2}{*}{Impact}} & Confidentiality ($C$) \\
    & \textbf{\multirow{2}{*}{Score ($\mathcal{I}$)}} & Integrity ($I$) \\
    & & Availability ($A$) \\ \hline
    & \multirow{2}{*}{\textbf{Temporal}} & Exploit Code Maturity ($E$) \\
    & \multirow{2}{*}{\textbf{Score ($\mathcal{T}$)}} & Remediation Level ($\mathit{RL}$) \\
    & & Report Confidence ($\mathit{RC}$) \\ \hline
    & \multirow{2}{*}{\textbf{Environmental}} & Security Requirements ($C_R$, $I_R$, $A_R$) \\
    & \multirow{2}{*}{\textbf{Score ($\mathcal{E}$)}}& Modified Base Metrics ($X_M$, where \\
    & & $X \in \{ \mathit{AV}, \mathit{AC}, \mathit{PR}, \mathit{UI}, S, C, I, A, \mathit{ISS}\}$) \\ \hline
\end{tabularx}

\medskip

\tcbsubtitle{\textbf{Other Definitions:}}

\medskip

\begin{tabularx}{\columnwidth}{rX}
$\mathit{ISS}$ & Original Impact Sub-Score. \\
$\mathcal{R}$ & Round up (i.e., \emph{ceil}) operation with scale of one.
\end{tabularx}

\medskip

\tcbsubtitle{\textbf{CVSS$_{v3.1}$ Base Score ($\mathbf{\mathcal{B}}$):}}

\begin{equation} \label{eq:cvss-impact-subscore}
    \mathit{ISS} = 1 - \big[\left(1-C\right) \times \left(1-I\right) \times \left(1-A\right)\big]
\end{equation}

\medskip

\begin{equation} \label{eq:cvss-base-impact}
    \mathcal{I} = 
    \begin{cases}
        6.42 \times \mathit{ISS} & \text{if } S = \overline{Changed}\\[4pt]
        [ 7.52 \times \left(\mathit{ISS} - 0.029\right) - & \text{if } S = Changed \\
        ~ ~3.25 \times \left(\mathit{ISS} - 0.02\right)^{15} ] & \\
    \end{cases}
    \end{equation}

\medskip

\begin{equation} \label{eq:cvss-base-exploitability}
    \mathcal{X} = 8.22 \times AV \times AC \times PR \times UI
\end{equation}

\medskip

\begin{equation} \label{eq:round}
    \mathcal{R}(n) = \lceil n \times 10 \rceil / 10
\end{equation}

\medskip

\begin{equation} \label{eq:cvss-base}
\mathcal{B} = 
    \begin{cases}
        0 & \text{if } \mathcal{I} \leq 0 \\[4pt]
        \mathcal{R}\left(\min\left[\left(\mathcal{I} + \mathcal{X}\right), 10\right]\right) & \text{if } \mathcal{I} > 0 \text{, } \\ & S = \overline{Changed}\\[4pt]
        \mathcal{R}\left(\min\left[1.08 \times \left(\mathcal{I} + \mathcal{X}\right), 10\right]\right) & \text{if } \mathcal{I} > 0 \text{, } \\ & S = Changed
    \end{cases}
\end{equation}

\medskip

\tcbsubtitle{\textbf{CVSS$_{v3.1}$ Temporal Score ($\mathbf{\mathcal{T}}$):}}

\medskip

\begin{equation} \label{eq:cvss-temporal}
    \mathcal{T}=\mathcal{R}(\mathcal{B} \times E \times \mathit{RL} \times \mathit{RC})
\end{equation} 

\medskip

\tcbsubtitle{\textbf{CVSS$_{v3.1}$ Environmental Score ($\mathbf{\mathcal{E}}$):}}
\begin{equation} \label{eq:cvss-environmental-iss}
\begin{split}
    & \mathit{ISS_M} = \min\big[ 1 - ((1 - C_R \times C_M) \times \\
    & ~ ~(1 - I_R \times I_M) \times (1 - A_R \times A_M)) , 0.915\big]
     \end{split}
\end{equation}

\medskip

\begin{equation} \label{eq:cvss-environmental-mi}
    \hspace{-3mm}\mathcal{I}_M = 
    \begin{cases}
        6.42 \times \mathit{ISS_M} & \text{if } S_M = \overline{Changed} 
        \\[4pt]
        \big[7.52 \times (\mathit{ISS_M} - 0.029) -  3.25 \times &  \text{if } S_M = Changed \\
        
        ~ ~(\mathit{ISS_M} \times 0.9731 - 0.02)^{13}\big] & \\
            
    \end{cases}
    \\
    \\
\end{equation}


\medskip

\begin{equation} \label{eq:cvss-environmental-me}
    \mathcal{X}_M = 8.22 \times \mathit{AV_M} \times \mathit{AC_M} \times \mathit{PR_M} \times \mathit{UI_M}
\end{equation}

\medskip

\begin{equation} \label{eq:cvss-ambiental}
    \hspace{-3mm}\mathcal{E} = 
    \begin{cases}
        \begin{split}
            &\mathcal{R}\big(\mathcal{R}(\min\big[(\mathcal{I}_M+\mathcal{X}_M), &\text{if }S_M=\overline{Changed}\\ 
            & ~ ~10\big])\times \mathit{RC} \times \mathit{RL} \times E \big)
            \\[4pt]
            &\mathcal{R}\big(\mathcal{R}(\min\big[1.08 \times(\mathcal{I}_M+\mathcal{X}_M),  &\text{if }S_M=Changed\\ 
            & ~ ~10\big])\times\mathit{RC}\times\mathit{RL}\times E\big) 
        \end{split}
    \end{cases}
\end{equation}

\smallskip 

\end{tcolorbox}
\caption{CVSS$_{v3.1}$ metrics overview and equations.}
\label{fig:overview_cvss31}
\end{figure}

The CVSS Temporal Score~$\mathcal{T}$ is derived from the Base Score $\mathcal{B}$ using Equation~\ref{eq:cvss-temporal} and incorporates factors prone to change over time: the state of exploit techniques (Exploit Code Maturity~$E$), the availability of remediation (Remediation Level~$RL$), and the confidence in the vulnerability description (Report Confidence~$RC$). 
Note that, although Equation~\ref{eq:cvss-temporal} is referred to as temporal, it does not imply a dependency on time \emph{per se}; rather, the Temporal Score~$\mathcal{T}$ changes only when experts reassess the vulnerability status and update the values of $E$, $RL$, or $RC$.
Noticeable, this temporal metric became the Threat metric group in CVSS$_{v4}$, which for clarity, we may use them interchangeably in many discussions in our paper.


The Environmental Score~$\mathcal{E}$ tailors the CVSS score to reflect the importance of the affected IT asset within its organisation. 
Similar to the Base Score~$\mathcal{B}$, $\mathcal{E}$ depends on the impact and exploitability of the vulnerability.
However, for~$\mathcal{E}$, these components are adjusted to fit the organisational context and are known as the modified impact~$\mathcal{I}_M$ and modified exploitability~$\mathcal{X}_M$. 
$\mathcal{I}_M$ is derived from Equation~\ref{eq:cvss-environmental-mi}, using the modified Impact Sub-Score~$\mathit{ISS}_M$ and the modified confidentiality, integrity, and availability metrics ($C_M$, $I_M$, and $A_M$), which are adaptations of those used in the standard impact calculation. 
Additionally, the security metrics $C_R$, $I_R$, and $A_R$ allow for further customisation, adjusting the CVSS score based on the criticality of the confidentiality, integrity, and availability of the organisation asset.
The modified exploitability~$\mathcal{X}_M$ uses Equation~\ref{eq:cvss-environmental-me}, which involves the modified Attack Vector~$AV_M$, Attack Complexity~$AC_M$, Privileges Required~$PR_M$, and User Interaction~$UI_M$. 
These metrics are customised by the organisation's analysts to reflect the specific environmental conditions surrounding the affected asset.

The Environmental Score~$\mathcal{E}$ is finally calculated according to Equation~\ref{eq:cvss-ambiental}, which incorporates the modified Scope metric ($S_M$) and integrates the temporal metrics---Exploit Code Maturity~$E$, Remediation Level~$RL$, and Report Confidence~$RC$---similarly to how they are used in the computation of the Temporal Score~$\mathcal{T}$ as per Equation~\ref{eq:cvss-temporal}. 
Following this CVSS$_{v3.1}$ overview, the discussion will shift to explore other existing scoring systems.

\section{Related Work} \label{sec:rel-wrks}

\subsection{Temporal and Dynamic Scoring in CVSS}

Although CVSS$_{v3.1}$ includes temporal and environmental scores, these metrics are not inherently time-dependent since they are only updated upon reevaluation by experts when significant changes in the vulnerability's status occur. 
This method, however, often fails to capture the dynamics of threat environments where the likelihood of exploitability can increase over time due to advancements in attacker capabilities and other external factors. 
CVSS$_{v4}$ renamed the temporal score to the threat metric group, which aims to provide a more continuous reflection of the current threat landscape. 
This transition underscores a critical shift from a predominantly reactive model to one that proactively adapts to emerging threats, enhancing the ability to prioritise vulnerabilities more effectively based on real-world risk assessments.
Following a similar observation, Minohara \emph{et al.}~\cite{minohara2023security} propose evolving the score of a vulnerability over time from its initial CVSS$_{v3.1}$ metrics (Base, Temporal, or Environmental) to a score augmented by the maximum Exploitability Score $\mathcal{X}$, assuming that exploitability is time-dependent only. 
However, their approach contrasts with the static nature of CVSS$_{v3.1}$ exploitability metrics ($AV$, $AC$, $PR$, $UI$), which typically remain unchanged. 
This method may also obscure critical differences among vulnerabilities by equalising scores over time, complicating prioritisation. 
Our work, \metric (detailed in Section~\ref{subsec:secscore}), counters theirs by integrating a real-world data-driven statistical model into CVSS, dynamically estimating the probability of exploit code development, offering a practical and evolving alternative to the static $E$ metric.

\subsection{ML Approaches for Vulnerability Scoring}
Recently, the application of machine learning (ML) to enhance vulnerability assessments has also gained traction. 
Sharma \emph{et al.}~\cite{sharma2023hybrid} propose a hybrid scoring system that recalibrates the weights of impact and exploitability scores in CVSS$_{v2}$, showing improvements over traditional static models where impact consistently outweighs exploitability. 
Kuhn \emph{et al.}~\cite{kuhn2023common} explores deep learning to automate the prediction of CVSS$_{v3}$ base metrics, aiming to expedite and improve the labour-intensive scoring process.
Albanese \emph{et al.}~\cite{albanese2023framework} and another study~\cite{9642043} explore models that assess the likelihood and exposure of vulnerable components and model attacker preferences, respectively, through advanced machine learning techniques. 
These approaches demonstrate a shift towards automating and enhancing the accuracy of vulnerability assessments~\cite{sharma2023hybrid, kuhn2023common, albanese2023framework, 9642043}.

Abuabed \emph{et al.}~\cite{abuabed2023stride}, on the other hand, focus on evaluating the feasibility of each attack by considering attack vectors, attack complexity, and authentication metrics through threat modelling. 
However, our research with \metric prioritises the exploitability of vulnerabilities to address them effectively. 
This focus is critical as only about $4\%$ of CVEs have public exploit code released within the first year of publication, and when such code appears, it is typically published within two days~\cite{USENIX_2020_Householder}. 
These observations underscore the urgency to prioritise and swiftly address critical vulnerabilities.

\subsection{Predictive and Exploitability Modeling}
Jacobs \textit{et al.}~\cite{Jacobs2020} introduced a model to prioritise vulnerabilities by predicting their exploit ``in the wild" instead of ``published exploits" in line with~\cite{Allodi2015}. 
The proposed model presented three practical strategies for vulnerability remediation: CVSS scores, published exploits, and vulnerability attributes. 
They evaluated each strategy using a rules-based approach and a machine learning model. 
Ultimately, they built a machine learning model with the obtained dataset. The proposal presented good accuracy, efficiency, and coverage results compared to CVSS.

The EPSS~\cite{epss} represents a significant evolution in vulnerability scoring by estimating the likelihood of exploitation using predictive models. 
Spanning three versions, EPSS has advanced from a basic model using 16 features in EPSS$_{v1}$ to a sophisticated XGBoost model analysing 1,477 variables in EPSS$_{v3}$. 
This progression includes expanding data sources to include real-time data from multiple security databases and social media, enhancing the system's responsiveness to new threats~\cite{Jacobs2021, jacobs2023enhancing}. 
Although EPSS has been improving its accuracy (from an F1-score of $0.299$ in EPSS~v1~\cite{Jacobs2021} to $0.451$ in v2~\cite{jacobs2023enhancing}, and $0.728$ in v3~\cite{epss}), concerns regarding input data integrity, model transparency for AI interpretability, seamless integration with CVSS, and adaptability to new trends persist~\cite{jacobs2023enhancing}.
Additionally, EPSS must be recomputed for all known vulnerabilities during each prioritisation round, which increases the required computational resources for the vulnerability management process.




\subsection{Statistical Relational Learning Approaches}

Unique methodologies to assess vulnerability exploitability involve statistical and relational learning. 
A notable study~\cite{10008055} employs the RDN-Boost algorithm within a Statistical Relational Learning (SRL) framework to analyse the relationships between products, vulnerabilities, and exploits, providing insights into the challenges of generic vulnerabilities. 
This approach underscores the complexity of accurately predicting exploit code associations and the need for tailored models to effectively address specific scenarios~\cite{10008055}.
However, estimating the probability of an exploit code or technique being related to a given vulnerability was beyond their scope.
 


\subsection{Emerging Trends and Open Challenges}

The field continues to evolve, with ongoing research highlighting the limitations of current methods and the necessity for continuous development. 
The dynamic nature of cyber threats necessitates that vulnerability scoring systems like CVSS adapt to changing conditions, enhancing their ability to predict and mitigate risks effectively. 
This is particularly critical as the context in which vulnerabilities exist can dramatically alter their impact and represent a greater (or lesser) risk over time~\cite{Howland2021}.
The present work considers the need to adapt the CVSS in order to better address the aforementioned constraints identified in the literature. To this end, it proposes an improvement in the computation of the CVSS Threat metric group (the equivalent to the Temporal metric in CVSS$_{v3.1}$) with empirical evidences.

\section{\metric}\label{subsec:secscore}

\metric is built on top of CVSS~\cite{cvss}, leveraging its strengths and adding mechanisms for better prioritisation of vulnerabilities in risk-based management processes. 
%
Although in this paper the discussion about \metric occurs mainly around CVSS$_{v3.1}$\footnote{Of the two most recent CVSS versions, CVSS$_{v3.1}$ has the most classifications available.}, its modularity also allows it to be applied to other CVSS versions. For this reason, in addition to detailing the \metric applied on CVSS$_{v3.1}$, this section also presents its formulation on CVSS$_{v4}$ as reference.

\begin{figure}[t!]
\centering
\begin{tcolorbox}[size=small,colback=A1,colframe=A3,title=,subtitle style={boxrule=0.4pt,colback=A3,colupper=black}]
\footnotesize
\vspace{-1mm}

\tcbsubtitle{\textbf{Definitions:}}
\smallskip

\begin{tabularx}{\columnwidth}{rX}
$\mathit{t}$ & Time (in weeks) since the CVE publication date. \\
$\mu$ & Location parameter of the AL distribution. \\
$\lambda$ & Scale parameter of the AL distribution. \\
$\kappa$ & Asymmetry parameter of the AL distribution. \\
$E_{min}$ & Minimum value that $E$ can assume in CVSS$_{v3.1}$. \\
$E_{max}$ & Maximum value that $E$ can assume in CVSS$_{v3.1}$. \\
\end{tabularx}

\medskip

\tcbsubtitle{\textbf{Asymetric Laplace (AL) CDF ($\mathbf{\mathit{F}}$):}}

\begin{equation} \label{eq:al_cdf}
    F(t;\ \mu,\lambda,\kappa)  = 
        \begin{cases}
          \dfrac{\kappa^2}{1+\kappa^2} \ e^{\tfrac{\lambda}{\kappa}(t-\mu)} & \text{if }t \leq \mu
          \\[8pt]
         1-\dfrac{1}{1+\kappa^2} \ e^{-\lambda\kappa(t-\mu)}  & \text{if }t > \mu
        \end{cases}
\end{equation}

\tcbsubtitle{\textbf{\metric Model for Exploit Maturity ($\mathbf{\mathit{E_\mathbb{S}}}$):}}

\begin{equation} \label{eq:SecScore_exploit_t}
    E_\mathbb{S}(t)=E_{min}+(E_{max}-E_{min}) \times F( t; \ \mu, \lambda, \kappa )
\end{equation} 



\tcbsubtitle{\textbf{\metric Temporal Metrics Group Score ($\mathbf{\mathit{\mathbb{S}}_\mathcal{T}}$):}}

\begin{equation} \label{eq:SecScore_temporal}
    \mathbb{S}_\mathcal{T}(t)=\mathcal{R}(\mathcal{B}\ \times RC\ \times RL) \times E_\mathbb{S}(t)
\end{equation} 

\tcbsubtitle{\textbf{\metric Environmental Metrics Group Score ($\mathbf{\mathit{\mathbb{S}}_\mathcal{E}}$):}}

\begin{equation} \label{eq:SecScore_ambiental}
    \hspace{-3mm}\mathbb{S}_\mathcal{E}(t)\!\!=\!\!
    \begin{cases}
        \begin{split}
            &\mathcal{R}(\mathcal{R}(\min\big[(\mathcal{I}_M\!+\!\mathcal{X}_M), 10\big])\times
            & \text{if }\!S_M\!=\!\overline{Changed}\\ 
            & ~ ~\mathit{RC} \times \mathit{RL}) \times E_\mathbb{S}(t)
            \\[4pt]
            &\mathcal{R}(\mathcal{R}(\min\big[1.08 \times (\mathcal{I}_M+ \mathcal{X}_M), & \text{if }\!S_M\!=\!Changed \\ 
            & ~ ~10\big]) \times RC \times RL) \times E_\mathbb{S}(t)
        \end{split}
    \end{cases}
\end{equation}

\end{tcolorbox}
\vspace{-3mm}
\caption{\metric metrics and equations applied to CVSS$_{v3.1}$.}
\label{fig:forensics}
\vspace{-2mm}
\end{figure}

\metric leverages a set of well-established CVSS metrics that are often quantified through careful expert assessments. The idea behind \metric is not to be an alternative to CVSS, but to complement it in order to address one of its main limitations for vulnerability prioritisation \cite{spring2021time,Howland2021, Spring2018, Tenable2020b, Tenable2020a, Tenable2020, Vulcan2019}, considering the real-world dynamics of exploit code development.  
Specifically, \metric converts the Exploit Code Maturity metric $E$ into a model $E_\mathbb{S}(t)$ that takes into account the probability of existing an exploit code to the vulnerability. 
Under this approach, \metric reassess the CVSS Temporal and Environmental Scores using a statistical model fitted to a dataset containing historical information of vulnerabilities. Furthermore, \metric is parameterised and can be customised for diverse vulnerability profiles, according to its type or platform.

As shown in Section~\ref{subsec:cvss}, although CVSS has a temporal metric $E$ to express the exploit code maturity, it is updated only when experts modify it based on perceived changes with respect to available exploit codes. 
This method of updating $E$ is limited, since the continuous efforts of hackers (white hat or black hat) can make exploit codes/techniques exist even if they are not publicly known. 
This means that the likelihood of having an unknown exploit code/technique for a given vulnerability tends to increase over time, and this effect is not expressed in $E$ as a function of time. 

The lack of time effect in the CVSS, specially regarding to $E$, may lead to an unoptimised or flawed prioritisation process. In a simple example, think about two vulnerabilities $V_1$ and $V_2$ that are published in the CVE Database in different dates ($V_1$ before $V_2$), both classified with the same CVSS score. 
At a given time $t_n$ after the publication of the two vulnerabilities, considering that the CVSS metrics remain unchanged, it would be difficult to discern which would be priority based on the CVSS score, given the way it is computed. 
However, it is reasonable to hypothesise that in $t_n$ the probability of having an exploit code/technique for $V_1$ tends to be greater than the probability of having an exploit code/technique for $V_2$, since $V_1$ tends to be older than $V_2$.

\begin{figure}[t!]
\centering
\begin{tcolorbox}[size=small,colback=A1,colframe=A3,title=,subtitle style={boxrule=0.4pt,colback=A3,colupper=black}]
\footnotesize
\vspace{-1mm}

\tcbsubtitle{\textbf{Definitions:}}
\smallskip

\begin{tabularx}{\columnwidth}{rX}
$EQ5$ & Macrovector for E with 3 levels: $\big[\{0,\mathit{E:A}\},\{1,E:P\},\{2,E:U\}\big]$\\
$E_{min}$ & Minimum value that $E$ can assume in CVSS$_{v4}$, where: \\
\end{tabularx}

\begin{equation} \label{eq:SecScore_CVSS4_min_e}
E_{min} = \frac{\mathbb{S}_{CVSS_{v4}} \big\rvert_{EQ5=2}}{\mathbb{S}_{CVSS_{v4}} \big\rvert_{EQ5=0}}
\end{equation} 

\begin{tabularx}{\columnwidth}{rX}
$E_{max}$ & Maximum value that $E$ can assume in CVSS$_{v4}$, where: \\
\end{tabularx}

\begin{equation} \label{eq:SecScore_CVSS4_max_e}
E_{max} = 1
\end{equation} 

\tcbsubtitle{\textbf{\metric Model for Exploit Maturity ($\mathbf{\mathit{E_\mathbb{S}}}$):}}

\begin{equation} \label{eq:SecScore_CVSS4_exploit_t}
    E_\mathbb{S}(t)=E_{min}+(E_{max}-E_{min}) \times F( t; \ \mu, \lambda, \kappa )
\end{equation} 

\tcbsubtitle{\textbf{\metric Threat Metrics Group Score ($\mathbf{\mathit{\mathbb{S}}_\mathcal{T}}$):}}

\begin{equation} \label{eq:SecScore_CVSS4}
    \mathbb{S}_\mathcal{T}(t)=\mathbb{S}_{CVSS_{v4}} \big\rvert_{EQ5=0}\ \times E_\mathbb{S}(t),
\end{equation}

\end{tcolorbox}
\vspace{-3mm}
\caption{\metric metrics and equations applied to CVSS$_{v4}$.}
\label{fig:overview_cvss4}
\vspace{-2mm}
\end{figure}

\metric aims to address this situation and others whose prioritisation can be better implemented by observing the effect of time on the probability of having an exploit code/technique for the vulnerability. 
In CVSS$_{v3.1}$, the approach of converting $E$ to $E_\mathbb{S}(t)$ is applied to CVSS's Temporal (\ref{eq:cvss-temporal}) and  Environmental (\ref{eq:cvss-ambiental}) equations, which are rewritten as (\ref{eq:SecScore_temporal}) and (\ref{eq:SecScore_ambiental}), respectively,
in which $\mathbb{S}_\mathcal{T}(t)$ and $\mathbb{S}_\mathcal{E}(t)$ are the Temporal and Environmental Scores provided by \metric, respectively. The computation of $E_\mathbb{S}(t)$ is done according to Equation~\ref{eq:SecScore_exploit_t},
wherein $E_{min}$ is the minimum value that $E$ can assume in CVSS ({\it i.e.}, 0.91 for the \texttt{Unproven} status in CVSS$_{v3.1}$), and $E_{max}$ is the maximum value that $E$ can assume in CVSS ({\it i.e.}, 1 for the \texttt{High} or \texttt{Not Defined} statuses in CVSS$_{v3.1}$). 

The asymmetric Laplace (AL) distribution is a generalisation of the Laplace distribution~\cite{kozubowski2000multivariate}, and its CDF is used in \metric to model the probability of existing an exploit code/technique for the vulnerability in a given time $t$ after its CVE date. 
The parameters $\{ \mu, \lambda, \kappa\}$ are adjusted to fit the CDF (\ref{eq:al_cdf}) to a vulnerability dataset. 
It should be noted that in \metric, the CDF (\ref{eq:al_cdf}) was chosen so that $\{ \mu, \lambda, \kappa\}$ can be adjusted and accurately represent vulnerabilities of different types, programming languages, platforms, among others. 
The performance of the AL CDF in representing the probability of existence of exploit codes is compared with the performance of other CDFs in Section~\ref{sec:results}.

Equation~\ref{eq:SecScore_exploit_t} converts the AL CDF $F( t; \ \mu, \lambda, \kappa )$ into an exploit code maturity metric $E_\mathbb{S}(t)$ that varies over the time from $E_{min}$ to $E_{max}$, as the probability of an existing exploit code/technique increases. 
Note that, in some cases the exploit code/technique may exist even before the official publication of the vulnerability in the CVE database~\cite{USENIX_2020_Householder, KDD_2019_Chen}, and this effect is supported by Equation~\ref{eq:SecScore_exploit_t} since Equation~\ref{eq:al_cdf} admits $F( t; \ \mu, \lambda, \kappa )>0$ for $t\leq0$.

Another aspect to highlight is the non-inclusion of $E_\mathbb{S}(t)$ in the rounding operation $\mathcal{R}$. 
The goal is to allow vulnerabilities to be distinguished even in cases where the difference between their \metric has a smaller scale than one decimal place. 
This approach considers the release of an exploit as a crucial factor for a prioritisation strategy and, when comparing different vulnerabilities, where even small differences in $E_\mathbb{S}(t)$ can indicate the greater or lesser probability of an exploit emerging for each one of them. 
Furthermore, applying $\mathcal{R}$ to $E_\mathbb{S}(t)$ could nullify its effect in distinguishing vulnerabilities for prioritisation, especially in low CVSS scores.

In addition to providing a continuous update of the vulnerability score based on the probability of having an exploit code for it over time, \metric also gives better treatment to vulnerabilities whose Exploit Code Maturity metric is classified as \texttt{Not Defined}. 
Note that, in all analysed CVSS versions~\cite{cvss,cvss4}, when the Exploit Code Maturity metric is classified as \texttt{Not Defined} then $E=1$, such as if it is classified as \texttt{High}, which is a pessimistic approach. 
A better way to handle these cases is especially relevant since, according to~\cite{vuldb} and shown in Figure~\ref{fig:vuldb_statistics}, the vast majority of vulnerabilities have $E$ classified as \texttt{Not Defined}. 
The problem is that by adopting this pessimistic strategy, the CVSS of most vulnerabilities (notably those classified as \texttt{Not Defined}) is raised to the same level as those in which $E$ is known to be \texttt{High}. 
Consequently, former vulnerabilities are prone to counter-productively dispute priority directly with the latter. 
In this sense, \metric can benefit most vulnerabilities ({\it i.e.}, those where $E$ is \texttt{Not Defined}) since, supported by statistical data, it does not immediately assign them the maximum $E$ score.

As previously mentioned, the \metric model is designed to be explainable and flexible to be applied to different CVSS versions. 
For this reason, we also present in Figure \ref{fig:overview_cvss4} the \metric applied to CVSS$_{v4}$, the most recent CVSS version although with fewer vulnerabilities classified than CVSS$_{v3.1}$. 
Note that the \metric Model for Exploit Maturity $E_\mathbb{S}(t)$, expressed by Equation \ref{eq:SecScore_CVSS4_exploit_t} in CVSS$_{v4}$, is the same as Equation \ref{eq:SecScore_exploit_t} in CVSS$_{v3.1}$. 
The value of $E_{max}=1$ is also the same in both cases.
The difference is in the definition of $E_{min}$, which in CVSS$_{v3.1}$ is tabulated \cite{cvss31}, but in the case of CVSS$_{v4}$ it is calculated by Equation \ref{eq:SecScore_CVSS4_min_e}, where $\mathbb{S}_{CVSS_{v4}} \big\rvert_{EQ5=0}$ and $\mathbb{S}_{CVSS_{v4}} \big\rvert_{EQ5=2}$ are the maximum and minimum CVSS$_{v4}$ scores that the vulnerability receives, respectively, according to what is attributed to the CVSS$_{v4}$ macrovector $EQ5$ (which represents the Exploit Maturity  \cite{cvss4}).
Finally, in Equation \ref{eq:SecScore_CVSS4}, $E_\mathbb{S}(t)$ is multiplied by $\mathbb{S}_{CVSS_{v4}} \big\rvert_{EQ5=0}$ so that $\mathbb{S}_\mathcal{T}(t)$ varies proportionally to the probability of existing an exploit code, as in Equations \ref{eq:SecScore_temporal} and \ref{eq:SecScore_ambiental} of CVSS$_{v3.1}$.


\section{Evaluation} \label{sec:results}

This section describes the dataset used to evaluate the \metric (\S\ref{sec:environment_datasets}), presents a comprehensive statistical analysis on publicly available vulnerability data (\S\ref{sec:secscore_statistics}), and demonstrates the effect of the \metric on vulnerability prioritisation (\S\ref{sec:impact_prioritisation}).

\subsection{Experimental Datasets} \label{sec:environment_datasets}


We downloaded snapshots (on April $2^{nd}$, 2023) of the following three well-known vulnerability databases and produced a dataset that complementarily aggregates their data:

\begin{itemize}
    \item \emph{CVE}~\cite{cve}, with more than $266$k CVE entries.\footnote{\url{https://cve.mitre.org/data/downloads/allitems.csv}.}
    \item \emph{ExploitDB}~\cite{ExploitDB-website}, with more than $45$k exploit entries.\footnote{\url{https://gitlab.com/exploit-database/exploitdb/-/blob/main/files_exploits.csv}.}
    \item \emph{CVEdetails}~\cite{cvedetails}, with the CVSS base score of more than $27$k CVEs.\footnote{\url{https://www.cvedetails.com/vulnerability-list.php}.}
\end{itemize}

The resulting dataset contains information about $27$k CVEs that are common to all three databases. The gathered data includes the CVE assignment date (between 2001 and 2022), a publicly known exploit (with its publication date), and a CVSS base score assigned.
It sizes approximately $2.6$MB and is publicly available on Zenodo\footnote{\url{https://zenodo.org/records/11123116}}.
The left part of Table~\ref{tab:ALD_param} details this resulting dataset, presenting the number of CVEs according to their types and affected platforms.

\subsection{Statistical Analysis on Publicly-Available Vulnerabilities} 
\label{sec:secscore_statistics}

As discussed in Section~\ref{subsec:secscore}, \metric reassesses the CVSS Score taking into account the probability of the existence of an exploit code/technique for a vulnerability over time. In \metric, such probability is modelled using the CDF of the asymmetric Laplace (AL) distribution, which can be adjusted to different vulnerability profiles by finding the appropriate $\{ \mu, \lambda, \kappa\}$ according to the characteristic of interest ({\it e.g.}, the vulnerability type, or platform). 
This section presents the results obtained with this \emph{data mining empirical approach~\cite{storey2020software}} and compares its accuracy with other statistical distributions, namely the Laplace (L) and the Skewed Normal (SN) distribution. All distributions were fitted to the dataset using the maximum likelihood estimation method.

To evaluate the ability of the \metric model to accurately represent different vulnerability profiles, the dataset was organised into three categories:
\begin{itemize}
    \item {\bf General}: which contains all vulnerabilities of the dataset;
    \item {\bf Type}: which organises the vulnerabilities into four subcategories, according to their types in~\cite{ExploitDB}: \texttt{DoS}; \texttt{Local}; \texttt{Remote}; and \texttt{Webapps}.
    \item {\bf Platform}: which organises the vulnerabilities into 24 subcategories, according to their platform classification in~\cite{ExploitDB}: \texttt{Android}; \texttt{iOS}; \texttt{Linux}; \texttt{macOS}; \texttt{Windows}; {\it etc}.
\end{itemize}

\begin{figure}[b!]
\centering\includegraphics[width=\columnwidth]{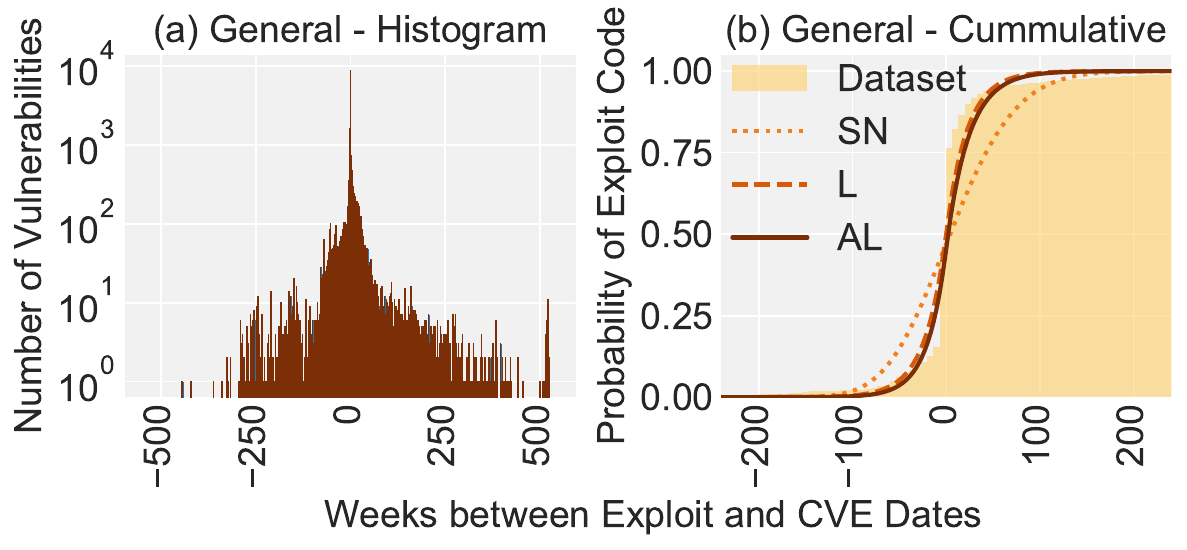}
\caption{Histogram of the time between the Exploit Code date and the CVE publication date.} \label{fig:exploit_code_maturity_stats}
\end{figure}

The profile of the General category is shown in Figure~\ref{fig:exploit_code_maturity_stats}, where the histogram presents the number of weeks elapsed from the CVE date to the Exploit date of all 27,432 vulnerabilities considered in this study. 
This figure provides a global view of the time it takes for a known vulnerability to have its exploit developed, based on the ExploitDB data. A negative number of weeks means that an exploit code/technique was available before the CVE date.

\begin{table*}[ht!]
\centering
\caption{The Dataset and the size of its subsets organised according to the CVE types and affected platforms; Mean Squared Error (MSE) for Skewed Normal (SN), Laplace (L), and Asymmetric Laplace (AL) distributions; Parameters of the AL distribution. Platforms with less than ten vulnerabilities were omitted.}
\label{tab:ALD_param}
\begin{tabularx}{\textwidth}{XXr|rrr|rrr}
\hline
\multicolumn{3}{c|}{\textbf{Datasets}} & \multicolumn{3}{c|}{\textbf{Mean Squared Error (MSE)}} & \multicolumn{3}{c}{\textbf{Parameters of the AL Distribution}} \\
\textbf{Category} & \textbf{Subcategory} & \textbf{Size} & \textbf{Skewed Normal} & \textbf{Laplace} & \textbf{Asym. Laplace} & $\mathbf{\mu}$ & $\mathbf{\lambda}$       & $\mathbf{\kappa}$ \\ 
\hline
General & \texttt{all} & $27432$ & \num{1.045e-04} & \num{5.539e-05} & \cellcolor{A1}\num{5.459e-05} & \num{-2.857e-01} & \num{2.179e+01} & \num{9.075e-01} \\
\hline
\multirow{4}{*}{Type} & \texttt{DoS} & $4022$ & \num{1.669e-04} & \num{1.106e-04} & \cellcolor{A1}\num{1.086e-04} & \num{1.429e-01} & \num{1.859e+01} & \num{9.493e-01} \\
 & \texttt{Local} & $2348$ & \num{8.441e-05} & \num{5.603e-05} & \cellcolor{A1}\num{5.258e-05} & \num{7.143e-01} & \num{3.210e+01} & \num{7.757e-01} \\
 & \texttt{Remote} & $4653$ & \num{6.494e-05} & \num{4.254e-05} & \cellcolor{A1}\num{4.078e-05} & \num{-3.840e-07} & \num{3.734e+01} & \num{7.513e-01} \\
 & \texttt{Webapps} & $16409$ & \num{2.937e-04} & \cellcolor{A1}\num{2.024e-04} & \num{2.027e-04} & \num{-4.286e-01} & \num{1.524e+01} & \num{1.071e+00} \\
\hline
\multirow{24}{*}{Platform} & \texttt{aix} & $38$ & \num{1.329e-04} & \num{1.141e-04} & \cellcolor{A1}\num{1.139e-04} & \num{5.143e+00} & \num{2.929e+01} & \num{9.655e-01} \\
 & \texttt{Android} & $109$ & \num{1.237e-04} & \num{1.208e-04} & \cellcolor{A1}\num{1.017e-04} & \num{7.143e-01} & \num{1.566e+01} & \num{4.828e-01} \\
 & \texttt{asp$^1$} & $1087$ & \num{3.810e-04} & \num{2.577e-04} & \cellcolor{A1}\num{2.532e-04} & \num{-4.286e-01} & \num{1.096e+01} & \num{1.229e+00} \\
 & \texttt{bsd$^2$} & $106$ & \num{1.914e-04} & \num{1.461e-04} & \cellcolor{A1}\num{1.437e-04} & \num{5.714e-01} & \num{2.938e+01} & \num{8.551e-01} \\
 & \texttt{cfm} & $42$ & \num{5.631e-04} & \num{5.041e-04} & \cellcolor{A1}\num{5.008e-04} & \num{-1.429e-01} & \num{3.108e+01} & \num{8.365e-01} \\
 & \texttt{cgi} & $445$ & \num{2.995e-04} & \num{2.562e-04} & \cellcolor{A1}\num{2.540e-04} & \num{-3.499e-09} & \num{2.694e+01} & \num{1.309e+00} \\
 & \texttt{hardware} & $1117$ & \num{1.429e-04} & \num{1.002e-04} & \cellcolor{A1}\num{9.306e-05} & \num{2.857e-01} & \num{1.878e+01} & \num{7.879e-01} \\
 & \texttt{HP-UX} & $19$ & \num{3.927e-05} & \cellcolor{A1}\num{3.406e-05} & \num{3.940e-05} & \num{7.000e+00} & \num{4.970e+01} & \num{2.012e+00} \\
 & \texttt{iOS} & $46$ & \num{4.638e-04} & \num{4.367e-04} & \cellcolor{A1}\num{3.855e-04} & \num{-1.000e+00} & \num{7.168e+00} & \num{1.750e-01} \\
 & \texttt{java} & $222$ & \num{1.215e-04} & \num{9.771e-05} & \cellcolor{A1}\num{8.519e-05} & \num{1.571e+00} & \num{1.680e+01} & \num{6.608e-01} \\
 & \texttt{jsp} & $269$ & \num{2.536e-04} & \num{1.801e-04} & \cellcolor{A1}\num{1.741e-04} & \num{-2.857e-01} & \num{1.338e+01} & \num{8.006e-01} \\
 & \texttt{Linux$^3$} & $2146$ & \num{1.002e-04} & \num{5.569e-05} & \cellcolor{A1}\num{5.431e-05} & \num{8.571e-01} & \num{2.559e+01} & \num{8.485e-01} \\
 & \texttt{macOS} & $80$ & \num{2.719e-03} & \cellcolor{A1}\num{2.674e-03} & \num{2.710e-03} & \num{1.414e+01} & \num{9.927e+00} & \num{8.082e-01} \\
 & \texttt{multiple} & $1927$ & \num{1.273e-04} & \num{8.930e-05} & \cellcolor{A1}\num{8.518e-05} & \num{1.857e+00} & \num{2.058e+01} & \num{8.733e-01} \\
 & \texttt{Novell} & $16$ & \num{7.719e-05} & \cellcolor{A1}\num{6.048e-05} & \num{6.387e-05} & \num{8.571e-01} & \num{1.657e+01} & \num{1.189e+00} \\
 & \texttt{osx} & $229$ & \num{1.694e-04} & \num{1.213e-04} & \cellcolor{A1}\num{1.005e-04} & \num{-1.429e-01} & \num{1.470e+01} & \num{6.365e-01} \\
 & \texttt{PHP} & $13261$ & \num{3.656e-04} & \num{2.662e-04} & \cellcolor{A1}\num{2.637e-04} & \num{-4.286e-01} & \num{1.456e+01} & \num{1.128e+00} \\
 & \texttt{python} & $21$ & \num{2.935e-03} & \num{2.775e-03} & \cellcolor{A1}\num{2.542e-03} & \num{5.472e-09} & \num{5.063e+00} & \num{3.694e-01} \\
 & \texttt{ruby} & $13$ & \num{9.859e-04} & \num{9.943e-04} & \cellcolor{A1}\num{9.561e-04} & \num{-5.286e+00} & \num{5.109e-03} & \num{1.085e-04} \\
 & \texttt{sco} & $12$ & \num{9.151e-04} & \num{8.974e-04} & \cellcolor{A1}\num{8.768e-04} & \num{1.400e+01} & \num{1.641e+01} & \num{2.143e+00} \\
 & \texttt{Solaris$^4$} & $87$ & \num{1.092e-04} & \num{9.905e-05} & \cellcolor{A1}\num{9.846e-05} & \num{3.143e+00} & \num{7.539e+01} & \num{9.213e-01} \\
 & \texttt{Unix} & $127$ & \num{8.597e-05} & \num{6.952e-05} & \cellcolor{A1}\num{6.939e-05} & \num{2.000e+00} & \num{5.345e+01} & \num{1.111e+00} \\
 & \texttt{Windows$^5$} & $5906$ & \num{6.039e-05} & \num{3.541e-05} & \cellcolor{A1}\num{3.269e-05} & \num{-1.429e-01} & \num{3.231e+01} & \num{7.502e-01} \\
 & \texttt{xml} & $55$ & \num{4.063e-04} & \num{3.691e-04} & \cellcolor{A1}\num{3.110e-04} & \num{-4.801e-09} & \num{6.180e+00} & \num{2.664e-01} \\
\hline
\multicolumn{9}{l}{$^1$ \texttt{asp = \{ashx, asp, aspx\}}; $^2$ \texttt{bsd = \{bsd, bsd\_x86, freebsd, freebsd\_x86, freebsd\_x86-64, netbsd\_x86, openbsd\}};} \\
\multicolumn{9}{l}{$^3$ \texttt{linux = \{linux, linux\_mips, linux\_x86, linux\_x86-64\}}; $^4$ \texttt{solaris = \{solaris, solaris\_sparc, solaris\_x86\}};} \\
\multicolumn{9}{l}{$^5$ \texttt{Windows = \{Windows, Windows\_x86, Windows\_x86-64\}}} \\
\end{tabularx}
\end{table*}

Figure~\ref{fig:exploit_code_maturity_stats} also presents the corresponding cumulative distribution of the General case, and graphically compares how the CDF of the AL, L and SN distributions fit to the complete dataset ({\it i.e.}, without separating vulnerabilities by type or platform). 
For a better assessment on how accurate these distributions represent the General case, it is computed the mean squared error (MSE) of the CDF of each distribution in relation to the dataset. Considering all vulnerabilities, the most accurate distribution is the AL with  $MSE_{AL}=\num{5.459e-05}$, followed by the L distribution with $MSE_{L}=\num{5.539e-05}$, and the SN distribution with $MSE_{SN}=\num{1.045e-04}$. The AL distribution parameters that provide this $MSE_{AL}$ are $\{\mu, \lambda, \kappa\}=\{\num{-2.857e-01},\ \num{2.179e+01},\ \num{9.075e-01}\}$.
In the General case, the slight better performance of the AL distribution over the L distribution is attributed to its asymmetry adjustment capability. In fact, when observing Figure~\ref{fig:exploit_code_maturity_stats}, it can be seen a slight asymmetry between the left and right sides of the histogram ({\it i.e.}, before and after the CVE date, respectively). 
This asymmetry expresses the existence of different dynamics in the development of exploits before and after the CVE date, suggesting that there is some degree of influence of the CVE publication on the development of exploits. 
Concerning the SN distribution, although its skewness property also allows it to be adjusted to asymmetric distributions, the results show that the AL distribution still fits better to the General dataset.

Recall that the \metric model was developed to be flexible and capable of representing different vulnerability profiles, with this customisation being achieved through the adjustment of the parameters $\{ m, \lambda, \kappa\}$ of the AL distribution. 
To demonstrate the accuracy and adjustability of \metric, the same analysis is carried out for the 28 different vulnerability profiles, which correspond to the aforementioned vulnerability subcategories: 4 different types of vulnerability; and 24 different platforms to which they belong.

Table~\ref{tab:ALD_param} compares the accuracy of the AL, L, and SN distributions on these 28 vulnerability profiles, providing the MSE obtained with each statistical model. 
The lowest MSE of each vulnerability profile (dataset subcategory) is highlighted. 
Note that in most cases the AL distribution presents the best performance ({\it i.e.}, the lowest MSE). In the best case, the MSE of the AL distribution is  \num{3.269e-05}, and in the worst case, it is \num{2.710e-03}, which indicates the good accuracy of the AL distribution when representing a wide range of vulnerability profiles. 
The only cases in which the L distribution slightly outperforms the AL distribution are on the vulnerabilities whose type is \texttt{WebApps} or the platforms are \texttt{HP-UX}, \texttt{macOS}, and \texttt{Novell}. 
Even so, it can be noted that in these cases where the L distribution performs better, the accuracy gap to the AL distribution is not significant, with the largest MSE difference being equal to \num{3.6e-05} in the case of \texttt{macOS}.

\begin{figure}[t!]
\centering\includegraphics[width=\columnwidth]{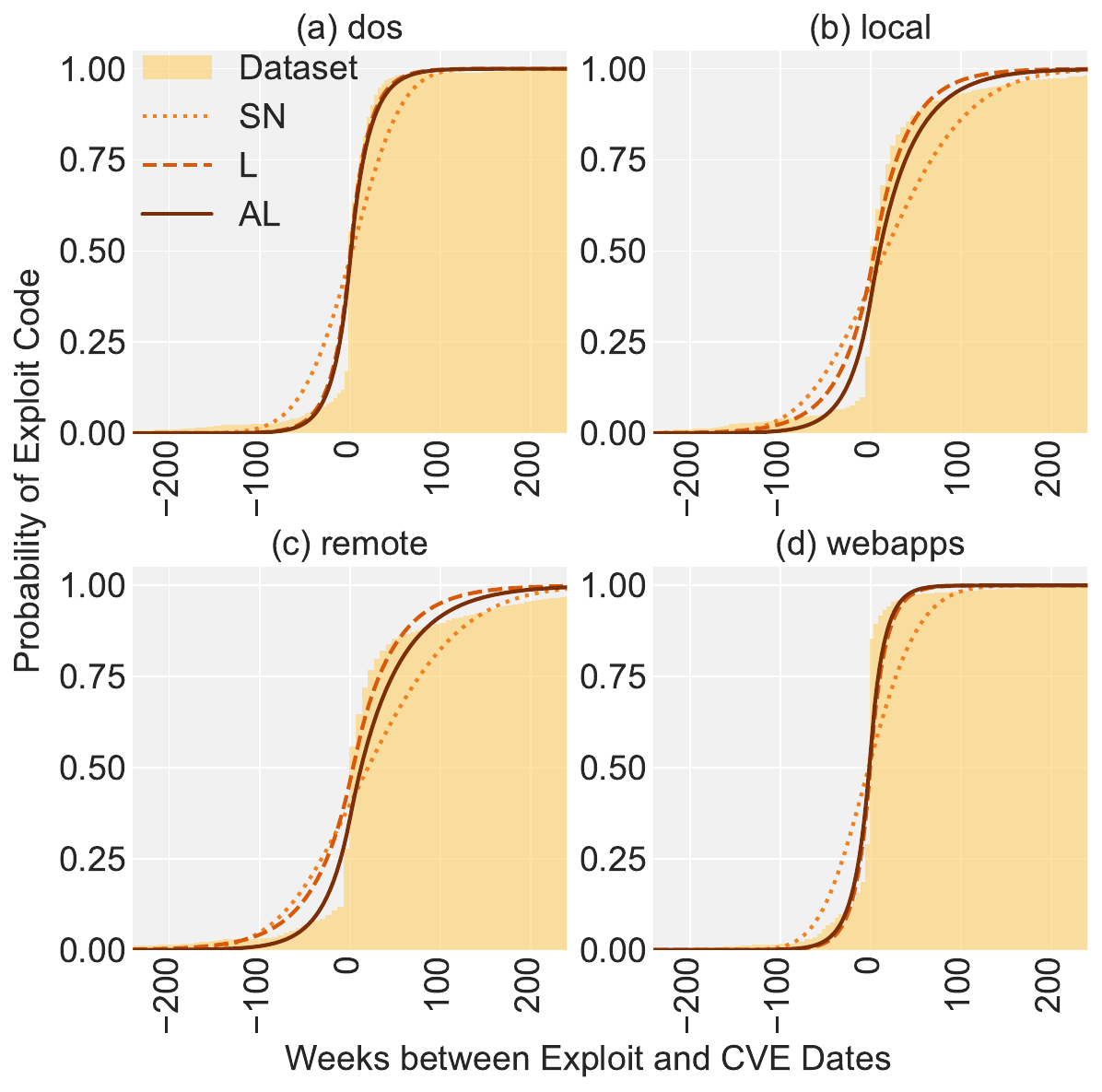}
\vspace{-2mm}
\caption{Cumulative distribution of exploit codes per vulnerability types.} \label{fig:types_vuln}
\vspace{-5mm}
\end{figure}

Note that in these cases, the worse performance of the AL distribution in relation to L can be attributed to the fitting algorithm, since the AL distribution admits a solution equal to that obtained by L.
This means that the CDF obtained with L can also be obtained with AL, incorporating the parameters of L into the AL distribution and assigning $\kappa = 1$.

Figure~\ref{fig:types_vuln} shows the probability of having an exploit code over time (taking the CVE date as reference) for the four different types of vulnerability, according to the dataset. 
This figure also presents the CDF of the SN, L, and AL distributions in order to allow a comparison of how each model fits the different exploit code probability profiles. 
In complement to the MSE data presented in Table~\ref{tab:ALD_param}, this figure graphically shows the accuracy of the AL distribution in representing the exploit code probability profile of these different types of vulnerabilities.

\begin{figure}[t!]
\centering\includegraphics[width=\columnwidth]{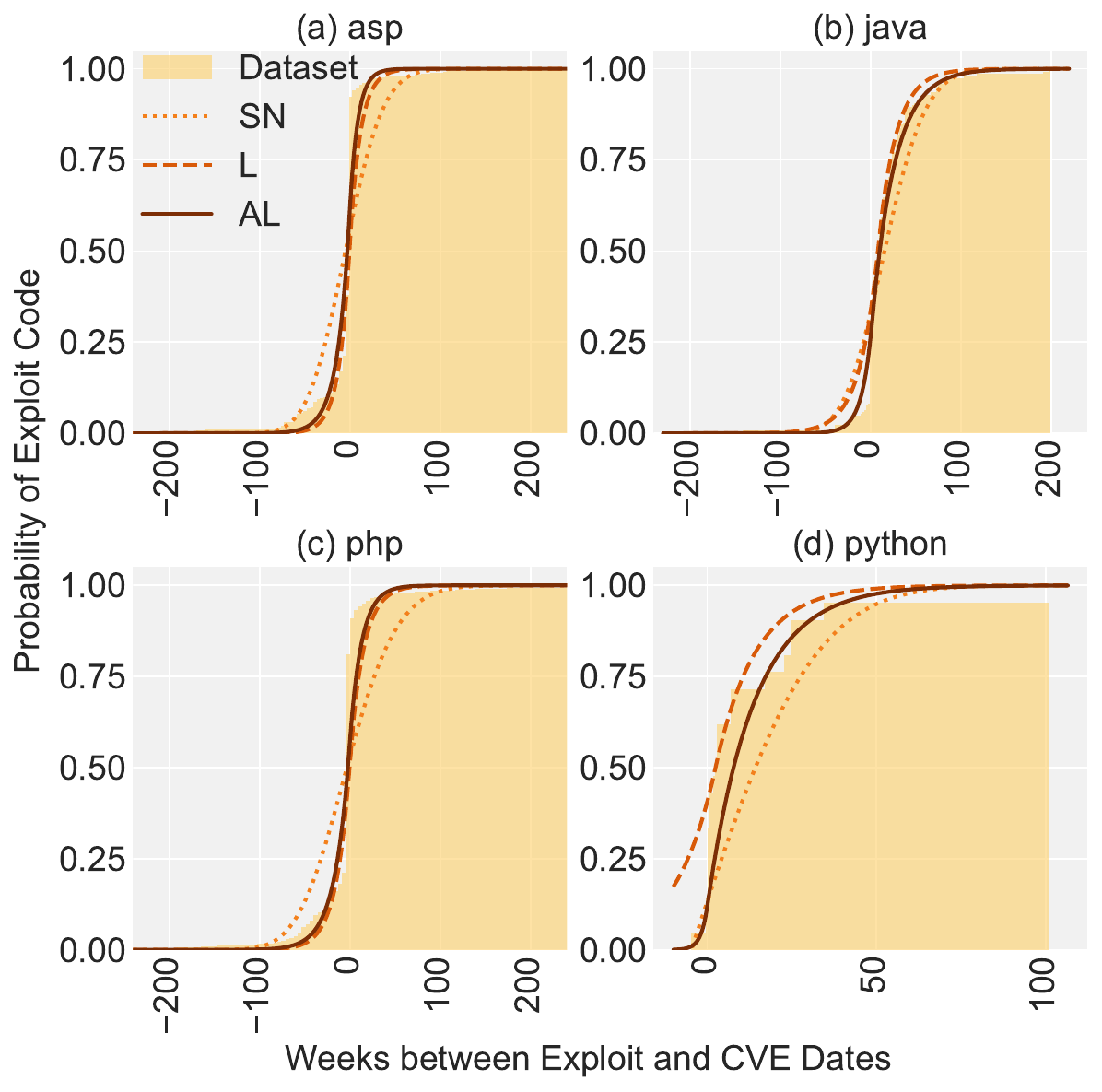}
\vspace{-2mm}
\caption{Cumulative distribution of exploit codes per programming language.} \label{fig:pl_vuln}
\vspace{-5mm}
\end{figure}

\begin{figure}[t!]
\centering\includegraphics[width=\columnwidth]{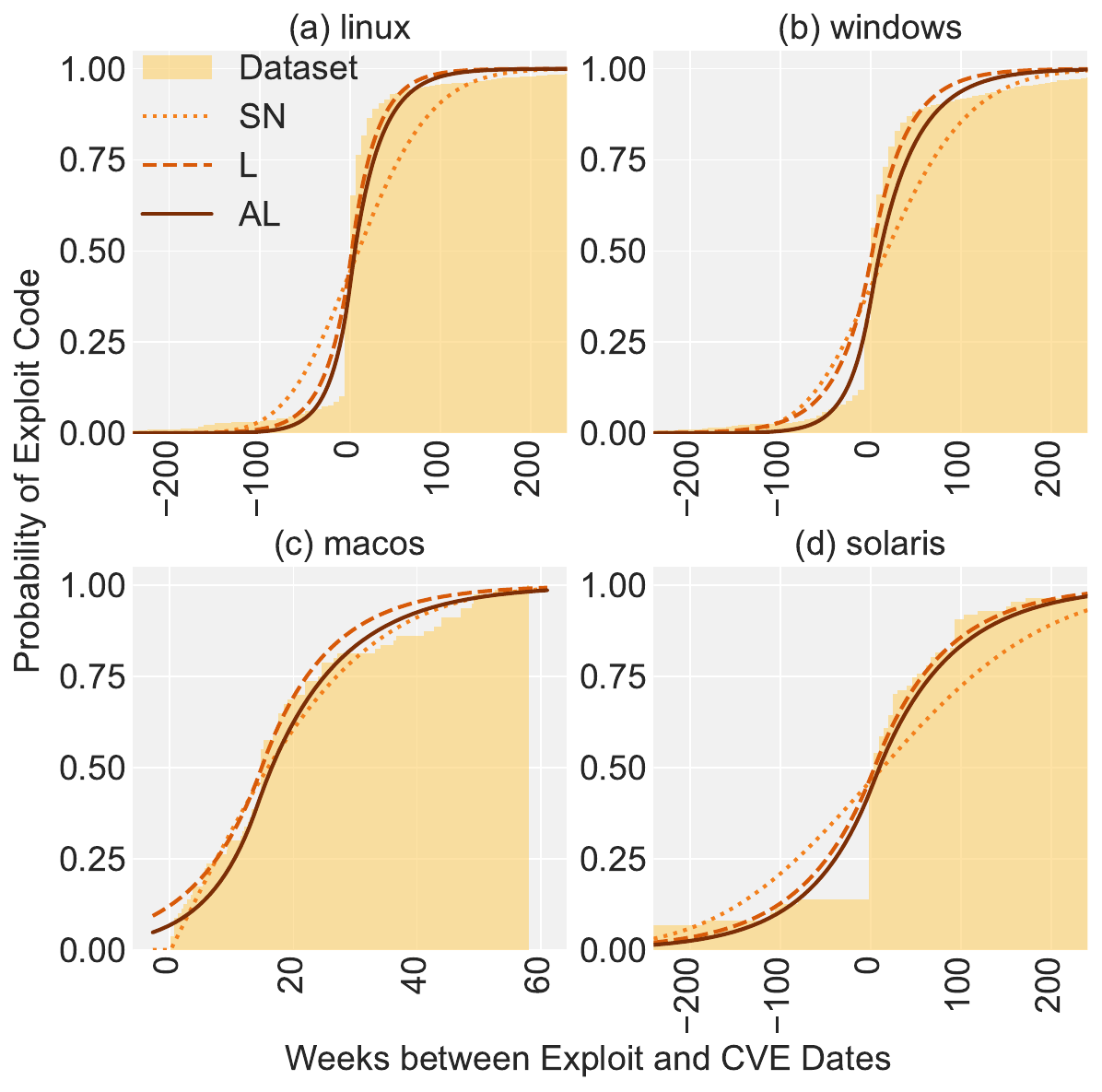}
\vspace{-5mm}
\caption{Cumulative distribution of exploit codes of OS-related vulnerabilities.} 
\label{fig:os_vuln}
\vspace{-5mm}
\end{figure}

\begin{figure}[t!]
\centering\includegraphics[width=\columnwidth]{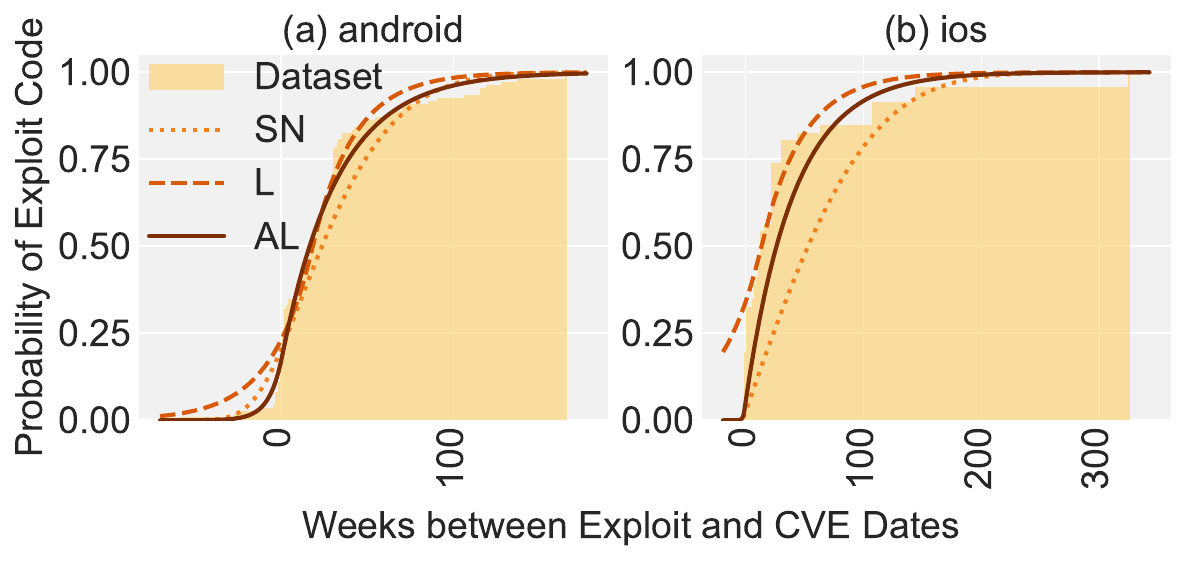}
\vspace{-7mm}
\caption{Cumulative distribution of exploit codes of vulnerabilities related to mobile OSes.} \label{fig:mobile_vuln}
\vspace{-6mm}
\end{figure}

The same representation is seen in Figure~\ref{fig:pl_vuln} for vulnerabilities organised according to examples of programming languages (namely \texttt{asp}, \texttt{java}, \texttt{php}, and \texttt{python}), showing the accuracy of the AL distribution in representing the exploit code probability profile of these vulnerability subcategories. 
More examples with similar observations can be found on the cumulative distributions of OS-related (Figure~\ref{fig:os_vuln}) and mobile-related (Figure~\ref{fig:mobile_vuln}) vulnerabilities.
The $\{ \mu, \lambda, \kappa\}$ values of the AL distribution of all vulnerability subcategories considered in this study are presented in Table~\ref{tab:ALD_param} for reference.

\subsection{\metric Effect on Vulnerability Prioritisation}\label{sec:impact_prioritisation}

This section demonstrates the effects of \metric on the prioritisation of vulnerabilities. All \metric values herein presented are computed based on the CVSS$_{v3.1}$ which, although less recent than CVSS$_{v4}$, still has more scores on databases available than the latter. However, it is worth highlighting that the \metric reasoning is also applicable to CVSS$_{v4}$, as discussed in Section~\ref{subsec:secscore}.

\begin{figure}[t!]
\centering\includegraphics[width=\columnwidth]{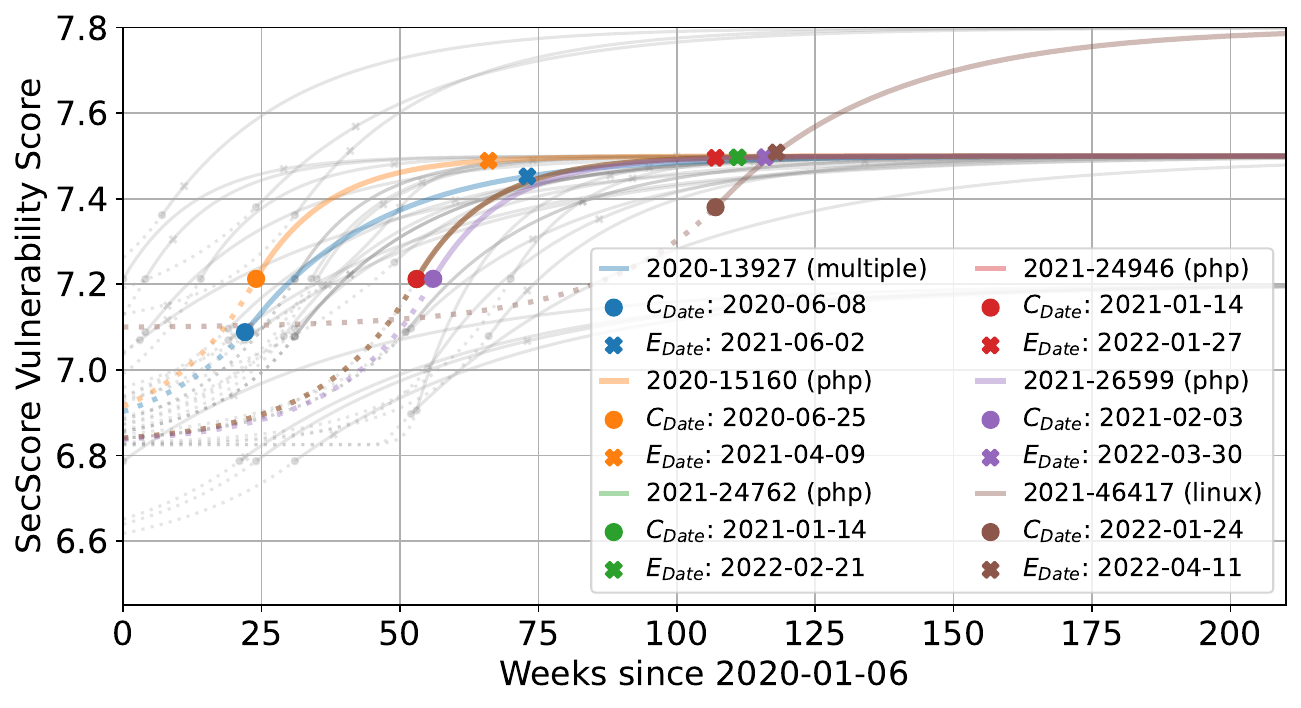}
\vspace{-7mm}
\caption{\metric evolution over time: vulnerabilities in different platforms published in 2020-2022 ($7\leq$ CVSS$_{v3.1}$ $\leq 8$).} \label{fig:2020_7_8_Platform}
\vspace{-5mm}
\end{figure}

To illustrate the effect of \metric on the vulnerability prioritisation process, we compare its evolution over time for different vulnerabilities. 
Figure~\ref{fig:2020_7_8_Platform} highlights six examples of vulnerabilities of different platforms, published between 2020 and 2021 with \mbox{$7\leq$ CVSS$_{v3.1}$ $\leq 8$}. 
The dashed and solid lines represent the \metric evolution before and after the vulnerability's CVE date ($C_{Date}$), respectively. 
The cross marks represent the date ($E_{Date}$) when the vulnerability's exploit code was published in \emph{ExploitDB}.

All vulnerabilities highlighted in Figure~\ref{fig:2020_7_8_Platform} received the same \mbox{CVSS$_{v3.1}=7.5$}, with the exception of CVE-2021-46417 which received a \mbox{CVSS$_{v3.1}=7.8$}. 
Note that, one could not rank them in a prioritisation process based only on their common CVSS of $7.5$.
However, some vulnerabilities may have the exploit code developed before others, as shown in Figure~\ref{fig:2020_7_8_Platform}, which the prioritisation process could take into account. 
Using \metric enables one to obtain different scores for these vulnerabilities and better prioritise them by taking into account the current time elapsed since the $C_{Date}$. 
This benefit can be verified in different examples contained in Figure~\ref{fig:2020_7_8_Platform}, as explained in the following remarks:

\paragraph{\textbf{(R1)}} Although the CVE-2020-15160 and CVE-2020-13927 vulnerabilities have close $C_{Date}$ and both have CVSS$_{v3.1}=7.5$, the \metric of CVE-2020-15160 remains higher than the \metric of CVE-2020-13927 over the time. 
This is because the probability of existing an exploit for \texttt{PHP} vulnerabilities tends to increase faster over time than for vulnerabilities of the \texttt{multiple} subcategory, according to the statistics presented in Section~\ref{sec:secscore_statistics}.
Note also that, even the $C_{Date}$ of CVE-2020-15160 being only 17 days after the $C_{Date}$ of CVE-2020-13927, the CVE-2020-15160 already has a higher \metric at its $C_{Date}$. 
The reason is that the probability of existing an exploit code for a \texttt{PHP} vulnerability at its $C_{Date}$ is greater than in the case of a \texttt{multiple} vulnerability, according to the statistics. 
The anticipated growth of such a probability is depicted by the dashed line of CVE-2020-15160. 
Following the time, it is possible to see that the $E_{Date}$ of CVE-2020-15160 indeed occurs before the $E_{Date}$ of CVE-2020-13927, which reinforces that CVE-2020-15160 must be prioritised as indicated by \metric. 
This decision would not be possible to conclude by considering only their CVSS.

\paragraph{\textbf{(R2)}} We can also compare the behaviour of \metric in an example with different vulnerabilities of the same platform and with the same CVSS$_{v3.1}$. 
This is the case of CVE-2020-15160, CVE-2021-24762, CVE-2021-24946, and CVE-2021-26599, which are all about \texttt{PHP}. 
Note that, although these vulnerabilities have different $C_{Date}$, they would all be levelled if analysed from the CVSS$_{v3.1}$ perspective only. 
However, we can observe that the $E_{Date}$ of these vulnerabilities ({\it i.e.}, the date they had an exploit code published in \emph{ExploitDB}) occur in the following sequence: CVE-2020-15160, [CVE-2021-24762, CVE-2021-24946], and CVE-2021-26599. 
This means that it would be better to prioritise them in this order, considering the sequence in which their exploits would appear. 
\metric allows vulnerabilities to be prioritised in advance ({\it i.e.}, even before the exploit emerges) based on the increased likelihood of an exploit being created. 
In fact, we can observe that the \metric of CVE-2020-15160 always remains above the \metric of [CVE-2021-24762, CVE-2021-24946] which, in turn, have their \metric always above the \metric of CVE-2021-26599. 
If, for example, a security analyst prioritise these vulnerabilities in week 60 of Figure~\ref{fig:2020_7_8_Platform}, \metric would consistently place them in that ordering thanks to its AL-based statistical model.
    
\paragraph{\textbf{(R3)}} \metric also allows vulnerabilities with different CVSS values to be consistently re-ranked and prioritised taking into account the likelihood of an exploit code emerging over the time. 
This is the case for CVE-2021-46417 (a \texttt{Linux} vulnerability with CVSS$_{v3.1}=7.8$) in relation to CVE-2021-24762, CVE-2021-24946, and CVE-2021-26599 (all \texttt{PHP} vulnerabilities with CVSS$_{v3.1}=7.5$).
Note that at $C_{Date}$ of CVE-2021-46417, and in the following weeks, the CVE-2021-46417 has a lower \metric than the other three. 
This means that based on the \metric, those three \texttt{PHP} vulnerabilities would have a higher priority than CVE-2021-46417, even though the latter has a higher CVSS value. 
In fact, the use of \metric in this prioritisation is advantageous since the $E_{Date}$ of the three \texttt{PHP} vulnerabilities indeed occur before the $E_{Date}$ of CVE-2021-46417, making reasonable to mitigate these \texttt{PHP} vulnerabilities first.

\begin{figure}[t]
\centering\includegraphics[width=\columnwidth]{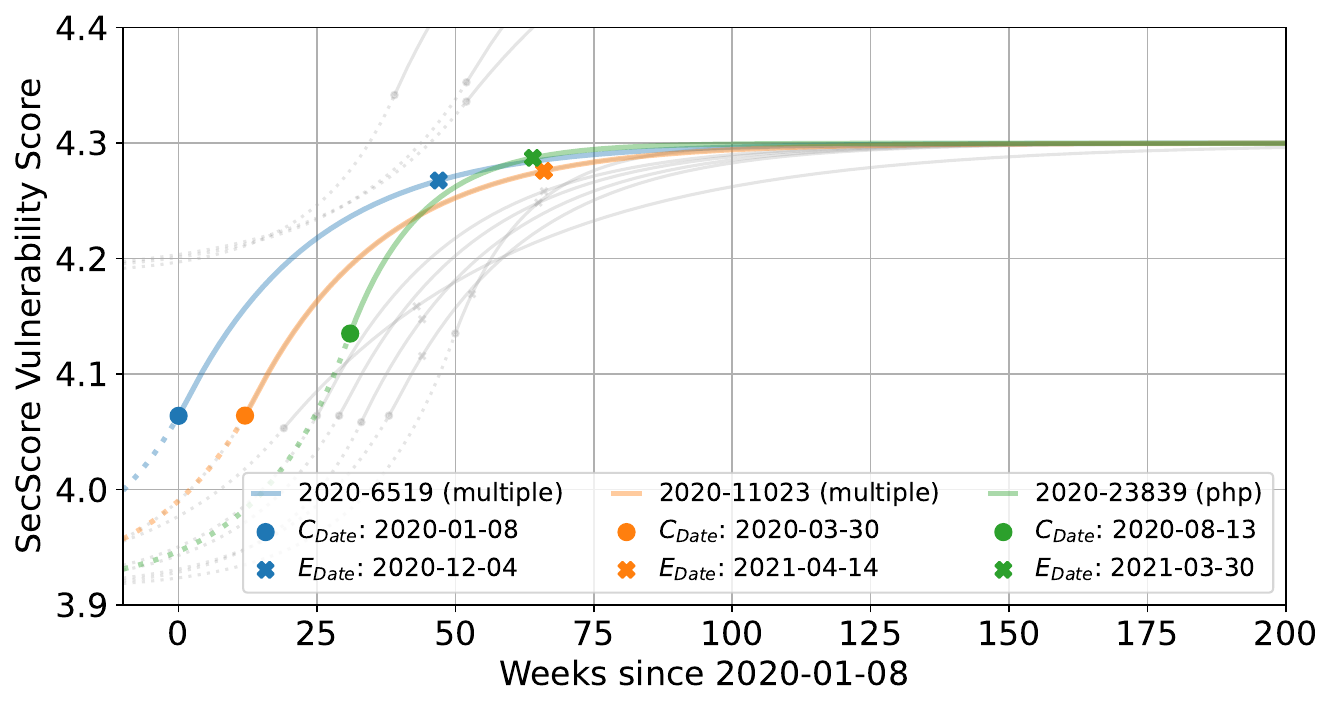}
\vspace{-7mm}
\caption{\metric evolution over time: vulnerabilities in different platforms published in 2020 ($4\leq$ CVSS$_{v3.1}$ $\leq 5$).} \label{fig:2020_4_5_Platform}
\vspace{-5mm}
\end{figure}

\vspace{2mm}
Figure~\ref{fig:2020_4_5_Platform} provides another example of how \metric addresses the effect of time in the prioritisation of vulnerabilities, by leveraging statistical data regarding the likelihood of existing an exploit code for each vulnerability. 
In this figure, three CVEs of different platforms (\texttt{multiple} and \texttt{PHP}) are highlighted as examples, all published in 2020 with CVSS$_{v3.1}=4.3$. 
The analysis of these examples is provided in the following remarks:

\paragraph{\textbf{(R4)}} CVE-2020-6519 and CVE-2020-11023 illustrate a case similar to that discussed in remark (R2). 
These vulnerabilities are from the same platform (\texttt{PHP}), but their CVE was disclosed on different dates, with the $C_{Date}$ of CVE-2020-6519 being an earlier date than that of CVE-2020-11023.
Considering that these vulnerabilities are from the same platform, they have the same statistical profile regarding the probability of an exploit code appearing. 
However, due to the earlier $C_{Date}$ of CVE-2020-6519, its \metric always remains above the \metric of CVE-2020-11023. 
This means that \metric suggests a greater prioritisation of CVE-2020-6519 than CVE-2020-11023 over time, which proves to be productive since the $E_{Date}$ of the former actually occurs before the $E_{Date}$ of the latter. 
This prioritisation would also not be possible based solely on the CVSS$_{v3.1}$ since both vulnerabilities would have the same score in that scoring system.

\paragraph{\textbf{(R5)}} In another case, one can note that CVE-2020-11023 has its $C_{Date}$ earlier than CVE-2020-23839 and, therefore, has a higher \metric when CVE-2020-23839 is published. However, over time, the \metric of CVE-2020-23839 ends up surpassing the \metric of CVE-2020-11023. This occurs due to two factors that derive from the statistics of \texttt{PHP} vulnerabilities. 
First, in its $C_{Date}$, CVE-2020-23839 has a greater probability of having an exploit code than CVE-2020-11023 had in its $C_{Date}$, reducing the \metric distance of both. 
Secondly, the probability of an exploit code emerging for \texttt{PHP} vulnerabilities grows faster than for vulnerability of the subcategory \texttt{multiple}, according to the AL-based statistical models obtained in this study. 
In fact, the increase in the \metric values of CVE-2020-23839 in relation to those of CVE-2020-11023 ends up providing better prioritisation, since the $E_{Date}$ of CVE-2020-23839 occurs before the $E_{Date}$ of CVE-2020-11023, as shown in Figure~\ref{fig:2020_4_5_Platform}. 
This is another example of prioritisation that would not be possible based only on the equal CVSS$_{v3.1}$ of these vulnerabilities, indicating the value and timeliness of \metric's statistical models to the prioritisation process.

These examples demonstrate the contribution of \metric to the vulnerability prioritisation process, but do not exhaust the cases in which this can be verified. 
\metric not only leverages the information provided by CVSS but also incorporates the time-dependent likelihood of existing an exploit code for the vulnerabilities. 
By making use of statistical data on exploit codes, \metric incorporates the time into the score computation, making it consistently change over time according to the vulnerability's profile.

Another important aspect to highlight is that the vast majority of vulnerabilities initially have their Exploit Code Maturity metric set to \texttt{Not Defined} ({\it i.e.}, $E=1$), and this classification tends to remain the same for many years, as seen in Figure~\ref{fig:vuldb_statistics}.
It means that these vulnerabilities receive the maximum score regarding this metric, being levelled with vulnerabilities whose Exploit Code Maturity is classified as \texttt{High}.
This can be counterproductive from a prioritisation perspective, since the vulnerabilities where $E$ is \texttt{Not Defined} in many cases an exploit code still does not exist. 
In \metric, the AL-based $E_\mathbb{S}(t)$ model presented in Figure~\ref{fig:forensics} causes the value of the Exploit Code Maturity to vary dynamically over time from the minimum value admitted by CVSS to its maximum ({\it i.e.}, from \texttt{Unproven} to \texttt{High}), in a manner proportional to the probability of the existence of an exploit code.
This approach gives fairer scores to vulnerabilities whose exploit code is classified as \texttt{Not Defined}, allowing them to gain priority as the probability of existing an exploit code increases (as shown in Figures~\ref{fig:2020_7_8_Platform} and~\ref{fig:2020_4_5_Platform}).
Evidently, if at any point an exploit code appears, $E_\mathbb{S}(t)$ needs to be set to its maximum value (and so will do \metric), regardless of the time elapsed, exactly how it is done reactively in CVSS nowadays.

It is worth mentioning that the results shown in this section do not take into account other variables often used in a risk-based management process. 
The aim here was to demonstrate how \metric can contribute to vulnerability prioritisation by leveraging statistical models built with empirical evidence from real-world exploit codes.
Notwithstanding, one can always integrate \metric with other works that use threat intelligence to specialise vulnerability metrics (e.g., adapting the CVSS Environmental metric group~\cite{hore2023towards}) for the organisational infrastructure context~\cite{Walkowski2022}.

\section{Conclusions} \label{sec:conclusions}


This paper presented \metric, an innovative vulnerability risk score that enhances the CVSS Threat metric group with empirical evidences for better vulnerability prioritisation.
\metric leverages the well stablished metrics and background from CVSS to complement it for vulnerability prioritisation in risk-based management processed.
\metric is more explainable than related works using machine learning and is flexible to be integrated into the prioritisation stage from multiple vulnerability management processed reviewed in Section~\ref{sec:bkg}.
Moreover, the design of \metric conducted to an empirical study into the likelihood of existing exploits for a vulnerability (see Section~\ref{sec:secscore_statistics}) and an experimental evaluation of \metric (see Section~\ref{sec:impact_prioritisation}), demonstrating timeliness for vulnerability prioritisation.

\subsection*{Acknowledgements}
This work was partially supported by the Digital European Programme under the ref.\ 101083770 and the Recovery and Resilience Plan (PRR) within the scope of the Recovery and Resilience Mechanism (MRR) of the EU, framed in the Next Generation EU within ATTRACT project, with ref.\ 774, and by the Fundação para a Ciência e Tecnologia (FCT) through the LASIGE Research Unit, ref.\ UIDB/00408/2020 (\url{https://doi.org/10.54499/UIDB/00408/2020}) and ref.\ UIDP/00408/2020 (\url{https://doi.org/10.54499/UIDP/00408/2020}).

\bibliographystyle{ACM-Reference-Format}
\bibliography{main}
\end{document}